\documentclass{wscpaperproc}
\usepackage{latexsym}
\usepackage{graphicx}
\usepackage{array}
\usepackage{multirow}
\usepackage{hhline}		
\usepackage{subfigure}
\usepackage{mathptmx}
\usepackage{mathtools}
\usepackage{amsmath}
\usepackage{amsfonts}
\usepackage{amssymb}
\usepackage{amsbsy}
\usepackage{amsthm}
\usepackage{todonotes}

\usepackage[pdftex,colorlinks=true,urlcolor=blue,citecolor=black,anchorcolor=black,linkcolor=black]{hyperref}

\newtheoremstyle{wsc}
{3pt}
{3pt}
{}
{}
{\bf}
{}
{.5em}
{}

\theoremstyle{wsc}

\begin{document}

%
%

\pagestyle{fancyplain}

\thispagestyle{plain}
\firstPageHead{}

\chead{\fancyplain{}{\itshape Hillmann, Uhlig, Dreo Rodosek, and Rose}}

\rhead{}
\cfoot{}
\renewcommand{\headrulewidth}{0pt} 

\makeatletter
\let\@internalcite\cite
\def\cite{\def\@citeseppen{-1000}%
    \def\@cite##1##2{(##1\if@tempswa , ##2\fi)}%
    \def\citeauthoryear##1##2##3{##1 ##3}\@internalcite}
\def\citeNP{\def\@citeseppen{-1000}%
    \def\@cite##1##2{##1\if@tempswa , ##2\fi}%
    \def\citeauthoryear##1##2##3{##1 ##3}\@internalcite}
\def\citeN{\def\@citeseppen{-1000}%
    \def\@cite##1##2{##1\if@tempswa, ##2)\else{}\fi}%
    \def\citeauthoryear##1##2##3{##1 (##3)}\@citedata}
\def\citeA{\def\@citeseppen{-1000}%
    \def\@cite##1##2{(##1\if@tempswa , ##2\fi)}%
    \def\citeauthoryear##1##2##3{##1}\@internalcite}
\def\citeANP{\def\@citeseppen{-1000}%
    \def\@cite##1##2{##1\if@tempswa , ##2\fi}%
    \def\citeauthoryear##1##2##3{##1}\@internalcite}
\def\shortcite{\def\@citeseppen{-1000}%
    \def\@cite##1##2{(##1\if@tempswa , ##2\fi)}%
    \def\citeauthoryear##1##2##3{##2 ##3}\@internalcite}
\def\shortciteNP{\def\@citeseppen{-1000}%
    \def\@cite##1##2{##1\if@tempswa , ##2\fi}%
    \def\citeauthoryear##1##2##3{##2 ##3}\@internalcite}
\def\shortciteN{\def\@citeseppen{-1000}%
    \def\@cite##1##2{##1\if@tempswa, ##2\else{}\fi}%
    \def\citeauthoryear##1##2##3{##2 (##3)}\@citedata}
\def\shortciteA{\def\@citeseppen{-1000}%
    \def\@cite##1##2{(##1\if@tempswa , ##2\fi)}%
    \def\citeauthoryear##1##2##3{##2}\@internalcite}
\def\shortciteANP{\def\@citeseppen{-1000}%
    \def\@cite##1##2{##1\if@tempswa , ##2\fi}%
    \def\citeauthoryear##1##2##3{##2}\@internalcite}
\def\citeyear{\def\@citeseppen{-1000}%
    \def\@cite##1##2{(##1\if@tempswa , ##2\fi)}%
    \def\citeauthoryear##1##2##3{##3}\@citedata}
\def\citeyearNP{\def\@citeseppen{-1000}%
    \def\@cite##1##2{##1\if@tempswa , ##2\fi}%
    \def\citeauthoryear##1##2##3{##3}\@citedata}
%
%
%
\def\@citedata{%
    \@ifnextchar [{\@tempswatrue\@citedatax}%
                  {\@tempswafalse\@citedatax[]}%
}

\def\@citedatax[#1]#2{%
\if@filesw\immediate\write\@auxout{\string\citation{#2}}\fi%
  \def\@citea{}\@cite{\@for\@citeb:=#2\do%
    {\@citea\def\@citea{, }\@ifundefined
       {b@\@citeb}{{\bf ?}%
       \@warning{Citation `\@citeb' on page \thepage \space undefined}}%
{\csname b@\@citeb\endcsname}}}{#1}}%

%
\def\@citex[#1]#2{%
\if@filesw\immediate\write\@auxout{\string\citation{#2}}\fi%
  \def\@citea{}\@cite{\@for\@citeb:=#2\do%
    {\@citea\def\@citea{, }\@ifundefined
       {b@\@citeb}{{\bf ?}%
       \@warning{Citation `\@citeb' on page \thepage \space undefined}}%
{\csname b@\@citeb\endcsname}}}{#1}}%

%
\def\@biblabel#1{}
\makeatother



\newdimen\bibindent
\bibindent=0.0em
\def\thebibliography#1{\section*{\refname}\list
   {}{\settowidth\labelwidth{[#1]}
   \leftmargin\parindent
   \itemindent -\parindent
   \listparindent \itemindent
   \itemsep 0pt
   \parsep 0pt}
   \def\newblock{}
   \sloppy
   \sfcode`\.=1000\relax}


\setlength{\baselineskip}{12.7pt}

\title{A Novel Multi-Agent System for Complex Scheduling Problems}

\author{Peter Hillmann\\
Tobias Uhlig\\
Gabi Dreo Rodosek\\
Oliver Rose\\
Department of Computer Science\\
Universit\"at der Bundeswehr M\"unchen\\
Neubiberg, 85577, GERMANY\\
}

\maketitle

\section*{ABSTRACT}

Complex scheduling problems require a large amount computation power and innovative 
solution methods. The objective of this paper is the conception and 
implementation of a multi-agent system that is applicable in various problem 
domains. Independent specialized agents handle small tasks, to reach a 
superordinate target. Effective coordination is therefore required to achieve 
productive cooperation. Role models and distributed artificial intelligence are 
employed to tackle the resulting challenges. We simulate a NP-hard scheduling problem 
to demonstrate the validity of our approach. In addition to the general agent 
based framework we propose new simulation-based optimization heuristics to given 
scheduling problems. Two of the described optimization algorithms are implemented using agents. This paper highlights the advantages of the 
agent-based approach, like the reduction in layout complexity, improved control 
of complicated systems, and extendability. 
\section{INTRODUCTION}

There is a rising interest in agent-based software development since it 
promises efficiency paired with a high mobility while approaching complex problems. 
Intelligent software agents are tools which handle their tasks independently 
within a larger context. These agents collectively operate in a multi-agent 
software (MAS) architecture, solving tasks by common endeavor. This paper 
introduces a concept of a highly flexible multi-agent system, that is applicable 
in various problem domains.

Our approach is evaluated using a combinatorial sequential scheduling problem 
and compared to traditionally object and component based approaches. We 
show, that using agents supports the development, by providing a better  
overview and thereby assist in handling the system complexity. With agents 
dependencies become visible and we obtain an impression at the phenomena of emergence.

Emergence refers to occurrence of new structures or properties, resulting from 
the cooperation of single elements in a complex system. These structures are not 
directly obvious and only arise from the interaction between the elements of the 
system \cite{UP10}. The common endeavor of more or less specialized directives 
surpasses the accumulated capability of all individual system elements 
\cite{Fer99}. The Micro-Macro-problem from the domain of distributed artificial 
intelligence illustrates this challenge very well \cite{Hil99}. Well known 
emergent systems are ant colonies and particle swarms. They have been adapted 
for optimization problems. 

There are two basic approaches to model the cooperation and organization in 
agent systems. We employ the role model which, in contrast to the service 
model, relies on describing the capabilities of an agent to perform a task 
instead of binding a fixed task to an agent.  Currently there is, however, no 
standardized concept for these architectures \shortcite{RWL95,Usbeck10}.

In Section 2 we describe the requirements we try to address using the MAS 
approach. Subsequently we provide a short overview of related work. In Section 
4, we present our MAS concept. Section 5 and 6 outline the application area and
the simulation-based optimization algorithms used for the 
evaluation in Section 7.
\section{REQUIREMENTS}

Agents are used in many different application areas, e.g., in information 
retrieval for automatic filtering, in electronic commerce for transactions 
\cite{XJ11}, in network management systems to detect intrusions \cite{Golling13}, 
or in simulations of entire hospitals \cite{Her07}. Usually a MAS is used when a 
single agent cannot meet all necessary requirements. Generally, the most 
important requirements are the following:

\begin{itemize}
\item Handle complex systems and challenges in a practicable way, relying on 
emergence \cite{BZW98}.
\item Guarantee flexibility by being extendible and adaptable, generally 
applicable, and platform independent. 
\item Provide a client-server environment for cluster systems enabling 
distributed calculation, parallelism, and dynamic load balancing.
\end{itemize}

\begin{figure}[htb]
{
\centering
\includegraphics[width=0.6\textwidth]{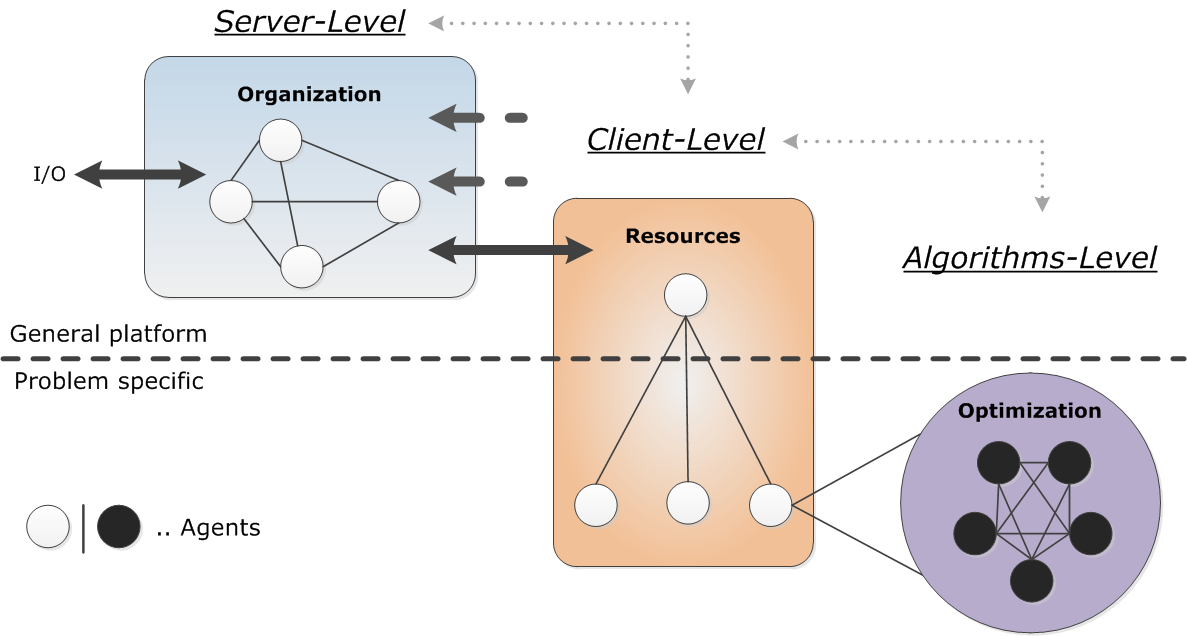}

\caption{Abstract architecture of the MAS.}
\label{Abstract}

}
\end{figure}

Figure \ref{Abstract} presents the overall concept of our approach relying on 
agents on different levels. The interaction of generally applicable and 
specialized agents guarantees an extendible approach.

\section{BACKGROUND AND RELATED WORK}
There is a plethora of agent architectures, however not single standard has 
emerged so far. The most common architecture is called reference model, which 
describes the general construction of a MAS. This reference is provided by the 
FIPA, a nonprofit group of various companies. This working group deals with the 
communication, interaction and management of agents as well as transport of 
agent messages and interoperability with different network architectures. Other 
well known agent architectures include: BDI-Modell \shortcite{Wooldridge99}, 
PAGE-Modell \cite{Russell09}, InteRRaP \shortcite{Oluyomi} and ACT-R 
\cite{Anderson96}. These architectures focus mainly on the description of 
agents, while providing only little information considering the cooperation and 
organization of multiple agents. A common but not necessarily optimal approach 
is using hierarchical structure to manage agents \cite{Sch00}. Consequently one 
agent controls multiple helper agents, which unfortunately often leads to the 
occurrence of bottlenecks. The realized system FABMAS \shortcite{Moench03} come close to our requirements and has strong influence on our design. But it has a hierarchical production control scheme. The system is also specific for semi conductor factories and maps their structure to the agent approach. 
The ideas in \cite{Yilamz09} describe a similar vision, but an implementation and feasibility evaluation are missing.
We want to promote a flexible MAS with independent 
agents, focusing on cooperation instead of simple delegation. It is therefor
mandatory that these intelligent agents exhibit autonomic behavior.

For the interaction and communication, the agents use the agent communication 
language (ACL) from FIPA \cite{Ode11b}. This language is very abstract and it 
relies on a intricate ontology service \shortcite{BCG07}. Based on the ACL, this 
paper presents a new concept to simplify and customize this aspect while keeping 
the flexibility.

\section{CONCEPT OF THE MAS}

Our concept relies on two types of agents. General applicable agents to realize 
the MAS and special agents with a problem dependent implementation. Instead of 
using a hierarchical structure we employ independent agents that cooperate using 
a kind of artificial intelligence. The MAS follows an model-view-controller 
principle.

The following section will illustrate the architecture of the MAS and describe 
the agents and its roles and functionality. Subsequently the agent 
communication and the processing of tasks is explained. Our agents register in a 
MAS by putting their name and role into a central listing-service. This is 
realized as a role of the agent and is an integral part of the environment. For 
the implementation, we use the framework JADE \cite{jade}. It offers a basic 
agent environments and agent functionalities. The simulation of the cluster tool 
is done using the SAGE framework. SAGE is an inhouse software solution  
including a discrete event simulator for typical scheduling problems and several 
optimization heuristics.
\subsection{Architecture of the MAS Lagoon}

\begin{figure}[htb]
{
\centering
\includegraphics[width=0.95\textwidth]{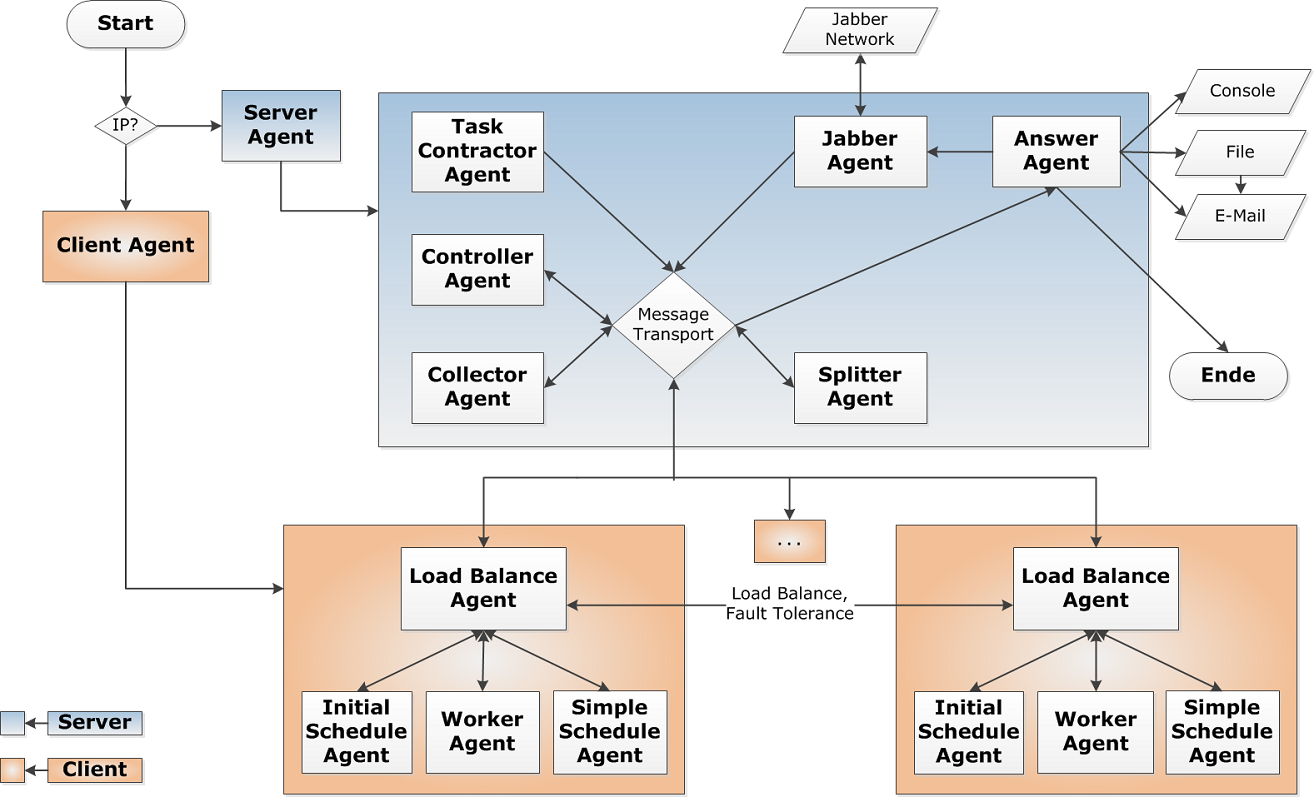}

\caption{Architecture of the MAS. The blue colored agents are part of the server 
application and the orange colored agents belong to the client version of the 
MAS. Communication interfaces are shown in rhomboids.}
\label{Lagoon}

}
\end{figure}

Our system is separated into a server and a client application to address the 
distributed processing (see Fig. \ref{Lagoon}). This enables the construction of 
a private or open cluster  combining the power of multiple host-systems for 
intensive computations.

Every agent we use relies on three basic behavior schemes. The 
\textbf{Initial-behavior} is run only once upon creation of an agent. The agent 
registers with his role at the listing-service and performs basic setup 
operations. After initialization the agent is either working or communicating. 
During \textbf{Working-behavior} the actual tasks are processed. The tasks 
arrive via messages. The \textbf{Messaging-behavior} relies on an incoming and an outgoing queue for messages. Each incoming task will be queued according to its priority. Control instructions are processed immediately. Other tasks are buffered for the Working-behavior. Outgoing messages will be send whenever the receiver is reachable and able to accept them. Working and messaging are processed in parallel with a separate thread for each behavior.

The constitution of the server part of the MAS starts with the creation of a 
\textbf{Server-Agent}. This agents initiates the MAS by spawning all server 
related agents. These are: Task-Contractor-Agent, Controller-Agent, 
Collector-Agent, Splitter-Agent, Answer-Agent, and Jabber-Agent. The Server-Agent is also 
responsible for eventually terminating the created agents.

The \textbf{Task-Contractor-Agent} loads the inital tasks, that should be 
processed by the MAS. Upon creation it loads a list of task and assigns 
priorities to them. Then it sends those task to appropriate Agents and 
terminates after sending all Tasks. The Task-Contractor-Agent is most basic 
option to provide tasks to the MAS.

The central manager for optimization intelligence is the  \textbf{Controller-Agent}. 
It chooses a fitting optimization approach and selects an adequate set of 
parameters. It adapts parameters dynamically during run-time and is capable of 
learning depending on the current state of the MAS.

The \textbf{Collector-Agent} takes a package of tasks and separates them into 
individual tasks for parallel processing. After completion of all tasks it 
collects them and evaluates the generated results. This offers the possibility to process a single task with different methods. It identifies appropriate 
solutions and transfers the results to other agents for further processing, e.g, 
output or learning. 

Parallel processing is facilitated by the \textbf{Splitter-Agent}. In contrast 
to the Collector-Agent which handles multiple tasks, it splits a single complex 
tasks into independent sub-tasks that can be processed in parallel. Ultimately 
the processed sub-task are reassembled to complete the original task. 

The output of the MAS is handled by the \textbf{Answer-Agent}, providing 
complete decoupling of input and output. The desired way of output communication 
is specified for each task and the Answer-Agent delegates the output to the 
appropriate interface/agent. Basic output is realized as writing to a file or 
sending a mail message. For more elaborate communication other agents are used.

For example the \textbf{Jabber-Agent} provides an interface to communicate using 
the instant messaging protocol XMPP \cite{XMPP} 
In addition to his role of providing output via XMPP the Jabber-Agent serves to accept 
input for the MAS. For example simple text messages can be used to start new 
tasks. 

The \textbf{Client-Agent} is similar to the Server Agent with regards to his 
main responsibilities but with a focus on the client side of the MAS. It 
individually generates agents to fulfill a certain role on the client. For 
example it could create a Load-Balancer Agents that is connected with the MAS 
supporting it with the available computation power on the client. Alternatively 
it creates an Contractor-Agent to submit tasks to the MAS. 

A key Agent is the \textbf{Load-Balancer-Agent} that manages the intersection 
between task management and actual task processing. Depending on the incoming 
tasks it dynamically creates specific worker-agents to handle them or assigns 
tasks to already existing workers. It is capable to manage the load of the 
client system, by creating an appropriate number of workers. Furthermore, it can 
dynamically delegate tasks to or request tasks from Load-Balancer-Agents 
operating on other clients. The Load-Balancer-Agent periodically exchange messages regarding their load state to identify agents with high or low load. This enable network wide load balancing in the whole 
MAS.

The group of \textbf{Worker-Agent}, as their name indicates, are the agents that perform 
the actual optimization. Depending on their setup they fulfill a certain role 
for processing. These Agents are highly problem related since they use a special 
algorithm to solve the given challenge. However we provide an general interface 
to easily implement a fitting worker-agent for each respective problem. With 
regard to our problem domain, the system has worker agents that generate initial 
schedules (Initial Schedule Agent) and other workers that iteratively optimize 
the initial schedule (Simple Schedule Agent). Idle Worker-Agents are terminated.

\subsection{Communication of the agents}

Communication in the MAS is realized using data packages based on the ACL (Agent 
Communication Language). Our messaging concept is simpler than the complete ACL 
architecture, but nevertheless remains universally applicable. We use wrapped 
objects that conform with the ACL, called \textbf{Coat-Package}. The idea is 
inspired by the OSI model and effectively decouples messaging in the MAS from 
the problem dependent parts in the tasks. A Coat-Package may transport any task 
and contains information not standardized by the ACL. Additionally to the specification it holds 
control parameters for the MAS, like priority or selected output channel. Each 
package has a unique identifier and a log for tracking purposes.

All information with problem dependent bits is kept in an \textbf{Task-Package}. 
For our scenario we use a Scheduling-Task-Package. It contains a test setup 
including all required parameters and the desired objective function. It also 
holds a list of necessary processing steps to generate a valid solution. A 
\textbf{Collection-Package} can be used to bundle multiple Task-Packages.

Problem independent \textbf{Control-Packages} enable the user to regulate the 
MAS during run time. Every agent has the required logic to process incoming 
control packages that should effect them. For example a control-package may be 
used to shutdown the MAS or activate debugging in agents. Valid targets for 
control packages are certain agents, all agents of the MAS or other packages.

During run time, the MAS is dynamic, e.g., client can join or leave the MAS. 
Therefore it is necessary to track all task, to avoid losing some of them. 
Managing agents keep a copy of a task until the receive a notice that a certain 
procedure has actually been complete by a worker agent. In case of failures, 
like connection losses, task are resubmitted to other available agents. 
Furthermore, it is possible to migrate agents from one system to another one 
during run time. This concept supported by the JADE framework guarantees failure 
safety based on redundancy. This approach is scalable and guarantees the redundancy to 
successfully operate in a dynamic environment.

\section{APPLICATION AREA}

We evaluate our approach applying it to a combinatorial sequence-dependent 
problem. This problem considers the scheduling of job for cluster tools in the 
production line of semiconductor manufacturing \cite{Unbehaun07}. Cluster tools 
can process jobs in parallel, however the processing time varies dependent on 
the processing of the jobs. Our objective is to minimize the completion time of 
the entire sequence to maximize the throughput. Figure \ref{fig: processing} 
illustrates the global improvement of processing jobs in parallel at the cost of 
slowing down the individual jobs due to shared resource. Discerning the optimal 
processing sequence correspond to the problem of finding the most fitting 
permutation of jobs with given recipes ($ R_{ i }$). The amount of permutations 
($S_{all}$) can be calculated with formula \ref{permutation}. Using Stirlings 
approximation illustrates the exponential growth of the resulting search space 
depending on the queue length (L).

\begin{figure}[htb]
{
\centering
\includegraphics[width=0.55\textwidth]{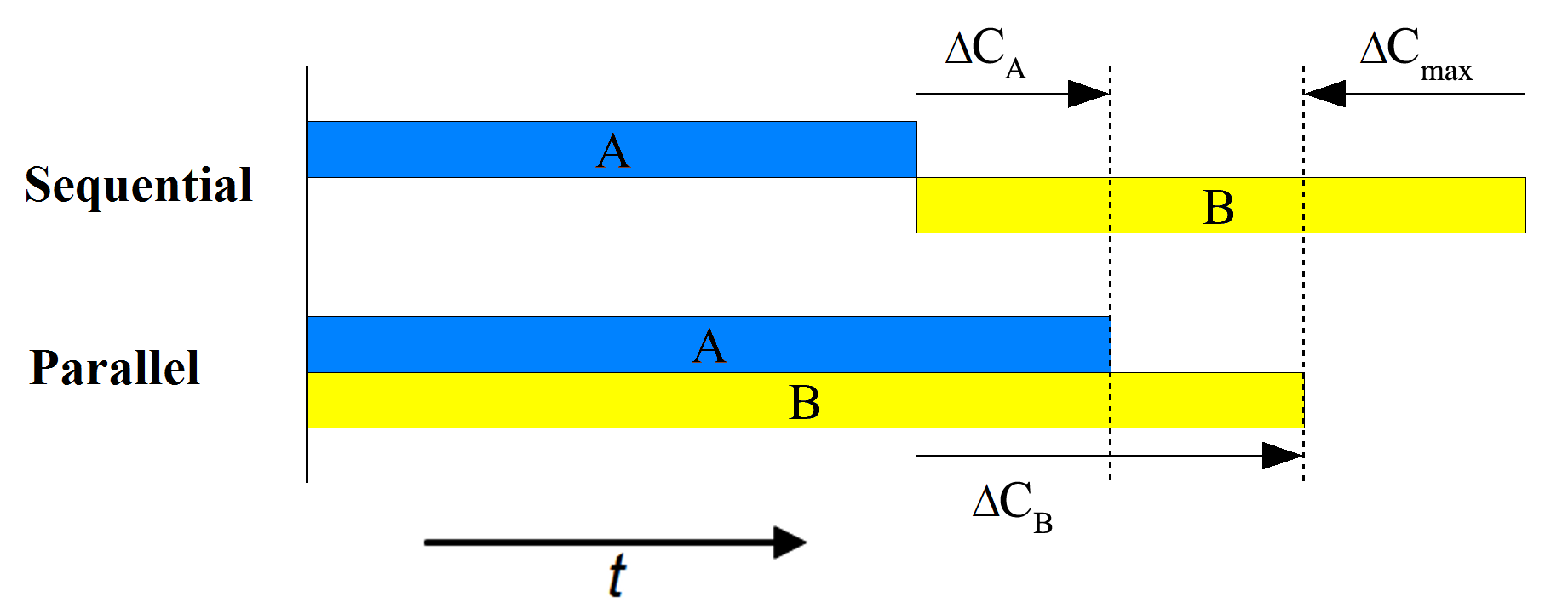}
\caption{Comparison of sequential and parallel processing.\label{fig: 
processing}}
}
\end{figure}

\begin{equation}
S_{all} = \frac{ L! }{ \prod^{ r }_{ i=1 }{ R_{ i }! } } = \frac{ L! }{ R_{ 1 }! \cdot R_{ 2 }! \cdot ... \cdot R_{ r }!}
\qquad
\qquad
L! \approx \sqrt{ 2 \cdot \pi \cdot L } \cdot \left(\frac{ L }{ e }\right)^{ L }
\label{permutation}
\end{equation}

The problem we face is a NP hard with a multidimensional and multimodal search 
space. An extended problem we consider employs parallel cluster tools, resulting 
a multi-step problem, which is still NP-hard, since the partitioning and 
sequencing influence each other.

\section{SCHEDULING ALGORITHMS}
\begin{figure}[htbp]
  \centering
  \begin{minipage}[b]{6,5 cm}
\includegraphics[width=\textwidth]{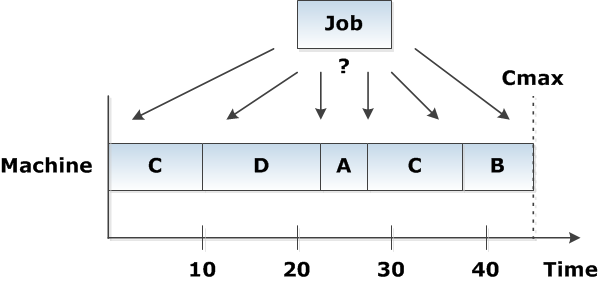}
\caption{Sequencing Problem.}
\label{fig: p1}
  \end{minipage}
  \qquad
  \begin{minipage}[b]{8,5 cm}
\includegraphics[width=\textwidth]{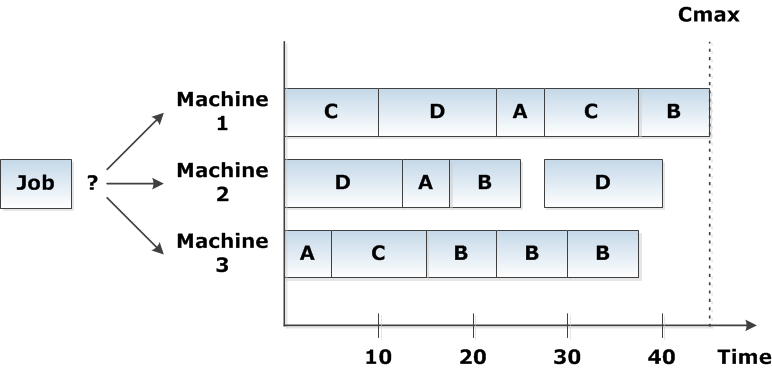}
\caption{Parallel Machine Problem.}
\label{fig: p2}
  \end{minipage}
\end{figure}

We developed different optimization algorithms for our reference problem. Of these algorithms, Particle-Swarm-Optimization and Central Complex are implemented as separate agent systems. Fundamentally the 
approaches fall into two categories -- sequencing and partitioning algorithms. 
Partitioning considers the assignment of jobs to a certain machine, while 
sequencing is about finding an adequate processing order of jobs on one machine 
(see Fig. \ref{fig: p1} and \ref{fig: p2}). Both, sequencing and partitioning, 
are performed in order to find a schedule that leads to a minimal makespan 
($C_{max}$). Both subproblems are NP-Hard and interlock with each other.

We consider two distinct approach to partition jobs. The basic approach attempts 
to discern an overall sequence of jobs, that is cut into equisized chunks. The 
resulting subsequences are assigned to the respective machines. The advanced 
approach is to use dynamic partitioning where the partitions are no longer 
require to be equisized. Dynamic partitioning is especially usefully for 
problems with multiple different machines and if you consider constraints like 
different processing speeds depending on the machine or machine qualifications 

\subsection{Random Down Swing}

The first proposed algorithm is a stochastic hill climbing approach with 
optional reinitialization. This Random Down Swing (RDS) approach is a simple 
sequencing algorithm -- for problems with multiple machines we uses the basic 
cutting approach to generate sequences. It starts with a randomized initial 
sequence and iteratively tries to improve the schedule by swapping two jobs in 
the sequence. The two jobs are selected randomly. A swap is accepted, if it leads to an improvement. The algorithm resets with a new initial sequence when a certain number of swaps were tried without improving the schedule. The limit before 
reinitialization is used, should be adapted to a given problem instance, for our 
scenarios we use an empirical found value of 700.  The optimization terminates 
after a fixed number of steps and returns the best found schedule.

\subsection{Simplex Method}
The Simplex Method is a popular algorithm for linear
programming and was developed by \cite{WM11}. It is part in the CPLEX
library from
IBM \cite{CPLEX10}.
This optimization algorithm is specialized for the search in constrained
solution spaces. The simplex method start at any vertices of constraints
and walk along edges of the constructed polytope to extreme points
with better objective values, until it reaches the optimum. It is based
on the assumption that the global optimum is part of the constructed
polytope, so the algorithms searches only on the edges of the
constraints. But the algorithms has performance problems with
multi-dimensional and multi-modal spaces, because of the high
variability. Other disadvantages are the adaption to the application
area as well as the worst-case complexity with exponential time
\cite{Klee72}.\\

\subsection{Particle-Swarm-Optimization}
The Particle-Swarm-Optimization (PSO) was originally proposed by \citeN{BS07} 
and is inspired by bird flocks. We adapted it to our problem and the discrete 
search space. Each particle is modeled as an individual intelligent agent. 
Therefore the resulting PSO is itself a MAS. Initially particles are spread 
randomly across the search space. We allow only positions that correspond to 
valid solutions and assign the respective fitness value to them. Each particle 
contains a sequence of jobs to encode a certain schedule. A single step to move 
in the search space is realized by swapping two jobs. With respect to the 
fundamental movement rules used in PSO, i.e., move towards global/local optimum 
or move randomly, we choose a suitable swap of two jobs for each particle.

Based on the agent approach, it is very simple to extend and improve the 
classical PSO. Additional agents are added to the algorithm as pseudo particle. 
These additional particle do not according to the classical PSO rules. These 
special particles encapsulate a traditional optimization heuristics, however the 
it can be influenced by the swarm and can in return influence the swarm.   For 
example the special agents may perform local optimization on solutions found by 
a moving particle and then provide the optimization results as guidance for 
other particles.

The MAS implementation of PSO has some interesting advantages. The main benefit 
is the asynchronous distributed optimization. Particle agents explore and 
evaluate solutions largely independently and therefor may be distributed in a 
network of multiple computers. Asynchronous exchange of information between 
agents allows a timely update of discovered global optima. Asynchronous 
communication can obviously lead to outdated information regarding the global 
optimum persisting in certain parts of particle swarm. We however observed no 
negative influence resulting from the obsolete communication artifacts.  

\subsection{Central Complex}
The Central Complex (CC) algorithm is a combination of partitioning and sequencing algorithms. The 
RDS algorithm is used to generate sequences for each 
machine. Partitioning is approached dynamically, starting from a random set of 
partitions. For each iteration RDS determines a good sequence for a given 
machine. Then a random job is shifted from the machine with the highest makespan 
to the machine with the lowest one. Shifting is managed in a way to avoid 
assigning jobs to unqualified machines. After shifting a job from one machine to another machine, the RDS optimizes the sequences again. Reinitialization is used to repeatedly 
start from different initial partitioning sets. A Meta-Optimization calculates 
appropriate parameters for the algorithm, e.g., number of reinitializations and 
iterations used for RDS. 

\section{SIMULATION AND ASSESSMENT}
We put our concept to the test, using ``Lagoon'' a prototypic implementation to optimize a set of scheduling problems. We chose a set of test setups reflecting typical real world challenges (see Table \ref{test-setups}). The performance of the algorithms stagnate at around 10k simulation calls.

\begin{table} [htbp]
\begin{center} 
\caption{Parameters of test setups. }
\label{test-setups}
\begin{tabular}{|l!{\vrule width 1.0pt}r|}
\hline
\multicolumn{2}{|c|}{Test Setups} \\
\hline
\hhline{--}
Machines & 1..4\\
\hline
Machine Types & uniform or mixed\\
\hline
Job Count & 16..60\\
\hline
Job Types & 3..10 recipes\\
\hline
Simulation Calls & 1k, 10k or 100k\\
\hline
Repetitions & 100\\
\hline
\end{tabular}
\end{center}
\end{table}

\begin{figure}[htbp]
  \centering
  \begin{minipage}[b]{8,0 cm}
    \centering
\includegraphics[width=1.00\textwidth]{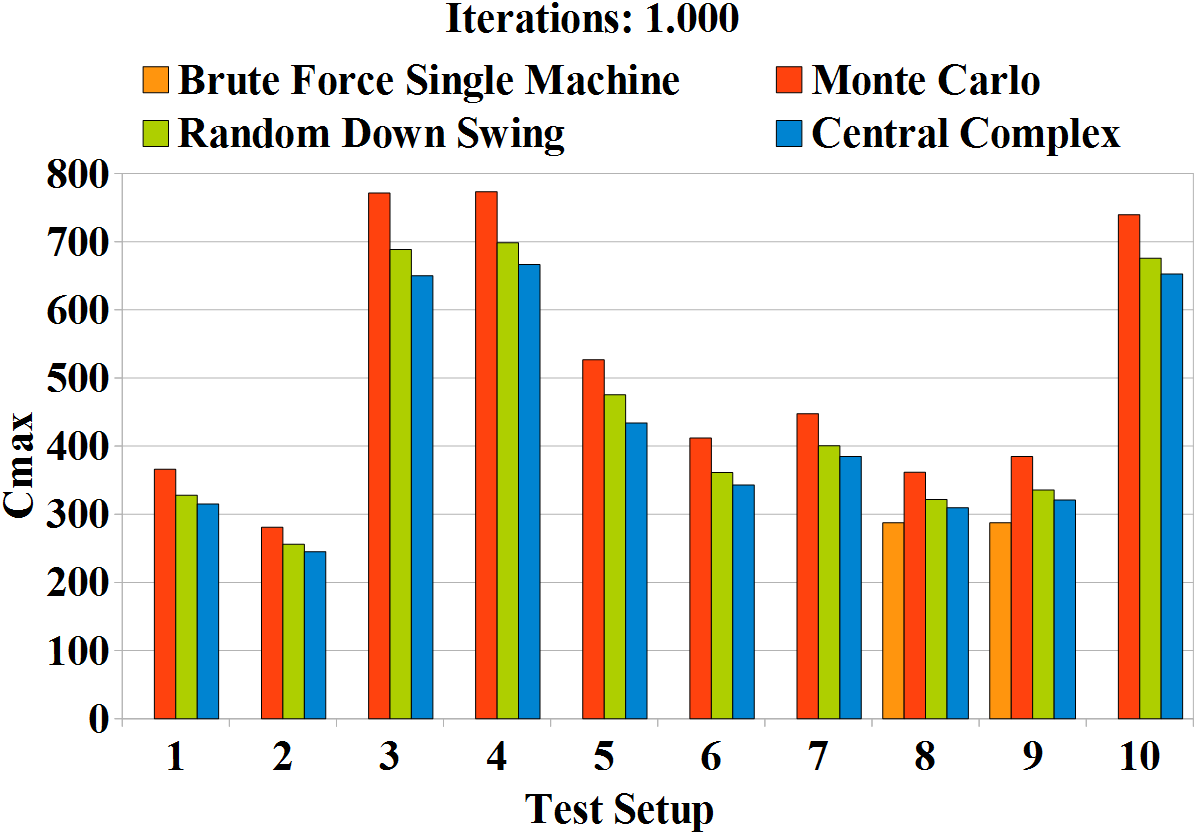}
\caption{Simulation results using 1.000 iterations. 
machines.\label{fig: sim11}}
  \end{minipage}
  \quad
  \begin{minipage}[b]{8,0 cm}
    \centering
\includegraphics[width=1.00\textwidth]{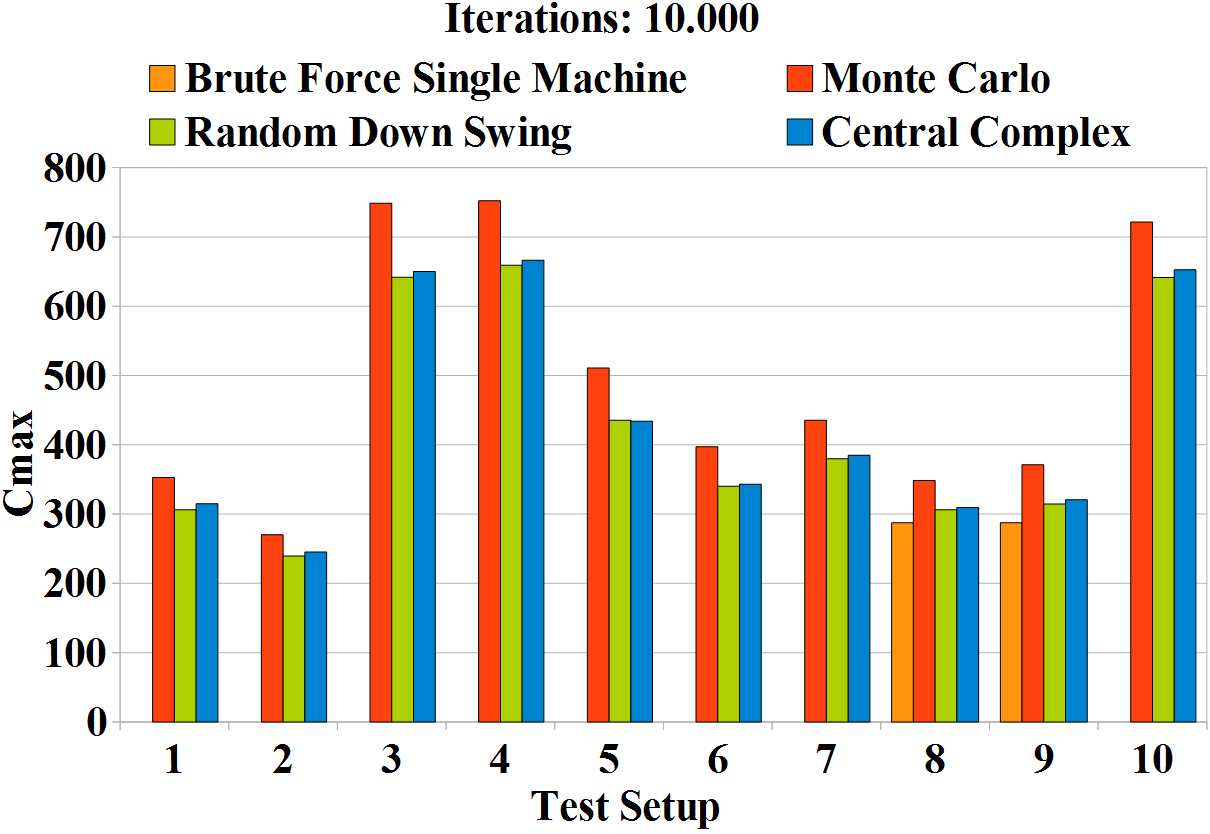}
\caption{Simulation results using 10.000 iterations. 
machines.\label{fig: sim12}}
  \end{minipage}
\end{figure}

Figure \ref{fig: sim11} and \ref{fig: sim12} summarize the results of our study. For two of the test problems we were able to calculate an optimal solution using a distributed brute force approach for an adapted single machine test setup. Generally a Monte Carlo (MC) optimization approach serves as a reference. It simply tests a random candidate solution for each iteration. An effective approach should obviously return better results than repeatedly guessing a solution.

As we expected our approaches exceeded the performance of MC. Random Down Swing 
(RDS) consistently provided the best results for all test setups when given enough calculation time 
(10k iterations). On the other hand the Central Complex method returned 
acceptable results with fewer iterations (1k). For test setup 5, a setup 
with a mixed set of machines and unequal quantity of job types, slightly outperformed RDS even for test using more 
iterations. Overall the quality of results was satisfying and came close to the 
optimum for the instances where brute force was used.

In addition to the numerical results we did observe the expected advantages of 
an MAS. It enabled us to effectively distribute the processing across multiple 
servers (mixed architecture containing N servers), significantly shortening the 
calculation time. During the development process we successively added new 
features benefiting from the flexibility and extendability of the approach. 

We conclude our survey with a comparison of our approaches in contrast to the simplex method. Table \ref{tab1} and \ref{tab2} show typical results for this comparison -- both of them are comparable in complexity. The theoretical value for the simplex method results from the calculated simplex tableau. In contrast the practical value is reached when we construct an actual solution based on the tableau. For test setup 2 the simplex method clearly outperformed the other approaches, returning a better overall result while needing less computation time. Regarding test setup 1 however, the simplex method fell short. The practical result did not come close to the predicted theoretical value and it took a relatively large amount of time to generate the solution. For this scenario even MC outperformed the simplex approach while the best results were achieved by RDS. 

\begin{table}[hbtp]
\centering
  \begin{minipage}[b]{7,8 cm}

\caption{Simulation of test setup 2}
\label{tab1}
\begin{tabular}{|l !{\vrule width 1.0pt} r|r|r|r|}
\hline
\multicolumn{1}{|l !{\vrule width 1.0pt}}{Algorithms} & \multicolumn{1}{l|}{Max} & \multicolumn{1}{l|}{Mean} & \multicolumn{1}{l|}{Min} & \multicolumn{1}{l|}{Time} \\
\hline
\hhline{-----}
Simplex-theo. 	& - 		& 371 & - 		& - \\ \hline
Simplex-prac. 	& - 		& 385 & - 		& 0.03 s\\ \hline
MC 		& 417 & 405 & 388 & 0.25 s\\ \hline
RDS				& 413 & 394 & 379 & 0.25 s\\ \hline
\end{tabular}

  \end{minipage}
  \qquad
  \begin{minipage}[b]{7,8 cm}

\caption{Simulation of test setup 1}
\label{tab2}
\begin{tabular}{|l !{\vrule width 1.0pt} r|r|r|r|}
\hline
\multicolumn{1}{|l !{\vrule width 1.0pt}}{Algorithms} & \multicolumn{1}{l|}{Max} & \multicolumn{1}{l|}{Mean} & \multicolumn{1}{l|}{Min} & \multicolumn{1}{l|}{Time} \\
\hline
\hhline{-----}
Simplex-theo. 	& - 		& 365 & - 		& - \\ \hline
Simplex-prac. 	& - 		& 421 & - 		& 26 s\\ \hline
MC 		& 436 & 416 & 396 & 0.26 s\\ \hline
RDS 					& 417 & 398 & 375 & 0.27 s\\ \hline
\end{tabular}

  \end{minipage}

\end{table}

\section{CONCLUSION AND OUTLOOK}

We reached a productive cooperation of agents without relying on a hierarchical 
structure or fixed grouping. The concept of role models supports the developer 
while organizing the intelligent agents. The combination of agents, role model and 
wrapped communication packages enables an orthogonal implementation of single 
components based on the model-view-control principle. Algorithms and other 
agents can be added to the MAS dynamically. Several users can send tasks to 
the MAS in parallel. We observed a reduction in layout complexity and improved 
control regarding complex systems. In contrast to traditional object-oriented 
and component-based approaches the agent concept is more flexible -- especially 
during runtime -- and supports distributed computation. Furthermore, the design of our MAS 
allows us to easily adapt it to various problem domains. This paper provides an 
approach that overcomes the biggest challenge of MAS, which is the need of an 
elaborate organization of agents.  

With regard to the scheduling domain our evaluation showed that computation 
intensive tasks can be solved efficiently using the developed MAS and the 
proposed simulation-based algorithms. Both RDS and CC algorithms reach reliably
good results.

\section*{ACKNOWLEDGMENTS}
This work was partly funded by FLAMINGO, a Network of Excellence project (ICT-318488) supported by the European Commision under its Seventh Framework Programm.\\

\appendix

\bibliographystyle{wsc}
\bibliography{demobib}

\section*{AUTHOR BIOGRAPHIES}
\noindent {\bf Peter Hillmann} is a Ph.D. student at the Universit\"at der Bundeswehr M\"unchen (UniBwM), Germany. He received his M.Sc. in Information-System-Technology in 2011 from Dresden University of Technology. His areas of research are network and system security with focus on cryptography as well as scheduling and optimization problems. His email address is \email{peter.hillmann@unibw.de}\\

\noindent {\bf Tobias Uhlig} is a research assistant at the Universit\"at der Bundeswehr M\"unchen, Germany. He received his M.Sc. degree in Computer Science from Dresden University of Technology. His research interests include evolutionary computation and its application to scheduling problems. He is a member of the IEEE RAS Technical Committee on Semiconductor Manufacturing Automation. His email address is \email{tobias.uhlig@unibw.de}\\

\noindent {\bf Prof. Gabi Dreo Rodosek} holds the Chair of Communication System and Network Security at the Universit\"at der Bundeswehr M\"unchen, Germany. She received her M.Sc. from the University of Maribora and her Ph.D. from the Ludwig-Maximilians University Munich. She is spokesperson of the Research Center Cyber Defence (CODE), which combines skills and activities of various institutes at the university, external organizations and the IT security industry (for instance Cassidian, IABG or Giesecke \& Devrient). Her email address is \email{gabi.dreo@unibw.de}.\\

\noindent {\bf Prof. Oliver Rose} holds the Chair for Modeling and Simulation at the Universit\"at der Bundeswehr M\"unchen, Germany. He received an M.Sc. degree in applied mathematics and a Ph.D. degree in computer science from W\"urzburg University, Germany. His research focuses on the operational modeling, analysis and material flow control of complex manufacturing facilities, in particular, semiconductor factories. He is a member of IEEE, INFORMS Simulation Society, ASIM, and GI, and General Chair of WSC 2012. His email address is \email{oliver.rose@unibw.de}.\\


\end{document}